    \documentstyle[nato,psfig]{crckapb}

\begin{opening}
\title{OBSERVATIONAL CONSTRAINTS TO THE EVOLUTION OF MASSIVE STARS}

\author{N. PANAGIA} 
\institute{ESA/Space Telescope Science Institute\\ 
3700 San Martin Drive, Baltimore, MD 21218, USA\\
E-mail {\it panagia@stsci.edu}}
\end{opening}

\runningtitle{Massive Star Evolution Constraints}

\begin{document}

\def\Te+{{$\rm{T_e}(\rm{O^+})$}}
\def\Te++{{$\rm{T_e}(\rm{O^{++}})$}}
\def\dZOdZ{{$\Delta$Z$_O$/$\Delta$Z~}}
\def\dYdZO{{$\Delta$Y/$\Delta$Z$_O$~}}
\def\dYdZ{{$\Delta$Y/$\Delta$Z~}}
\def\Msun{{M$_{\odot}$~}}
\def\msun{{\rm{M}_{\odot}~}}

\newcommand{\kms}{$km~s^{-1}$ }
\newcommand{\eg}{{\it e.g., }} 
\newcommand{\ie}{{\it i.e.~}} 
\newcommand{\etal}{{\it et al.}} 
\newcommand{\apj}{ApJ, } 
\newcommand{\aap}{A\&A, }
\newcommand{\apjl}{ApJ, } 
\newcommand{\aj}{AJ, } 
\newcommand{\mnras}{MNRAS, }

% The \begin{document} command comes after the \end{opening}
% command.

\begin{abstract}
We consider some aspects of the evolution of massive stars which can
only be elucidated by means of ``indirect" observations, \ie
measurements of the effects of massive stars on their environments. We
discuss in detail the early evolution of massive stars formed in high
metallicity regions as inferred from studies of HII regions in
external galaxies.

\end{abstract} 

\vskip -.5cm 

\section{Introduction} 

Massive stars play a crucial role in the evolution of galaxies and the
whole Universe, because they are the primary sources of radiative
ionization and heating of the diffuse medium, they provide most of the
nucleosynthesis products to boost the metal content of galaxies and the
intergalactic medium, and they constitute a major supply of kinetic
energy for galaxies, both through stellar winds during their quiescent
phases and, eventually, in the form of fast ejecta from supernova
explosions. 

Therefore, it is fundamental to reach a proper understanding of the
formation processes, the detailed properties and the evolution of
massive stars.  Despite the fact that their high luminosities make
them ``easy" targets for detailed observational studies, many aspects
and properties of massive star evolution are far from being fully
understood. This is because in any stellar generation, massive stars
constitute a small fraction of the newly formed stars (say, less than
1\% by number), they are ``elusive" in that their lifetimes are very
short (say, less than 10--20 million years), and often they are
heavily obscured by the parent molecular clouds where they were
formed, making even their identification rather cumbersome.  As a
consequence it is not easy to cover all evolutionary phases with {\it
direct} observations of a statistically significant sample of objects.

One can overcome these difficulties and gain additional insights by
considering phenomena that {\it indirectly} can provide hints and
clues to the problem. In other words, besides studying individual
massive stars, one can look at the effects that these stars have on
their environments (\eg HII regions, circumstellar nebulae, SNRs), and
infer from there what the stars were doing in special phases of their
evolution (\eg formation, LBV and pre-SN phases, etc.) that would not
be accessible in other ways. 

Thus, one can use radio observations of supernovae, which probe the
circumstellar material ejected by the progenitor stars several
thousand years before explosion, to study the very last phases of
their evolution. These phases represent a tiny fraction of a massive
star lifetime, $\sim$0.1\%, and, therefore, they are extremely
difficult to reveal and study with direct observations.   Although
this is an interesting aspect, we are not going to review it here, but
rather we refer the reader to recent papers (Montes \etal~ 1998,
Panagia \etal~ 1999, Weiler \etal~ 1999). 

Here, we consider and discuss one  particular aspect of the evolution
of massive stars, namely their formation and early evolution in high
metallicity environments.  We will show that observations of HII
regions in external galaxies show that the ionization of He is much
lower than that of H when the O/H ratio in the gas is appreciably
higher than solar. This implies that at high metallicities either very
massive stars ($M>25~M_\odot$) do not form, or they never reach their
expected ZAMS location. 

\section{Ionized Helium in the Milky Way}

There is clear observational evidence in the H~II regions of our
Galaxy that the fractional abundance of ionized helium
n(He$^+$)/n(H$^+$) is not a monotonic function of the galactocentric
radius. Moving outwards from the Galactic Center, the ionized He
abundance is found to increase in the inner Galaxy, then it attains a
maximum near the solar circle, and finally drops in the outer Galaxy
(\eg Mezger \& Wink 1983 and references therein). The negative
gradient in the outer galaxy reflects a genuine decrease in the He
abundance in the outward direction (\eg Panagia 1980; G\"usten \&
Mezger 1982). The positive gradient in the inner Galaxy instead is an
effect of the radial metallicity gradient which produces a systematic
variation of the spectrum of the ionizing radiation (Panagia 1980).

The fractional ionization of helium is extremely sensitive to the most
energetic part of the radiation field powering an H~II complex. Hence,
it can provide valuable information on the presence and the abundance
of the most massive ($m \ge$ 20 \Msun) stars, which are responsible
for most of the radiation with energy in excess of 24.6~eV. Therefore,
it is a powerful tool to study how the details of the star formation
process vary in different physical environments. There are several
mechanisms through which a higher metallicity lowers the He ionization
in an H~II region: \\ 
$\bullet~$ The relative number of He-ionizing photons in the stellar 
spectrum is reduced because of both a stronger line blanketing in the 
200--500 \AA ~wavelength range, and a higher continuum opacity.\\
$\bullet~$ The stellar radius becomes larger and the effective temperature 
decreases for a star of given mass, because of the increased continuum 
opacity in the sub-atmospheric layers of the star.\\
$\bullet~$ The upper cut-off of the Initial Mass Function (IMF, $m_U$, 
may be shifted 
to lower masses (\eg Kahn 1974; Shields \& Tinsley 1976).\\
$\bullet~$ A higher metallicity may induce a steeper IMF (i.e. a larger value 
of the slope $\alpha$ of the IMF $N(m) \propto m^{-\alpha}$) at least 
for $m >$ 10 \Msun, where the bulk of the ionizing radiation is produced
(\eg Terlevich \& Melnick 1983).

In the first two cases, metallicity acts ``directly'' on the radiation
field of the ionizing star cluster, by modifying the stellar spectra
without affecting the star formation processes. In the third and
fourth case instead, metallicity acts ``indirectly'' and the changes
in the radiation field result from changes in the properties of the
IMF. 

Panagia (1980) demonstrated that the combined effects of at least
the first three processes are needed to explain the He ionization in
the Milky Way. Moreover, these processes appear to account for the
observed gradient of the effective temperature of the ionizing
radiation inferred from the fitting of theoretical models to
observations of low-metallicity objects (Talent 1980; Campbell 1988).

\section{Ionized Helium in External Galaxies } 

Considering external galaxies, several authors (\eg Pagel 1986, 
Viallefond 1988, Robledo-Rella \& Firmani 1991) have suggested that a
systematic change of the IMF with metallicity is required by observations.
Others (\eg Fierro, Torres - Peimbert \& Peimbert 1986, McGaugh 1991) have
come to the opposite conclusion, and the controversy is still open.
A thorough assessment of this subject is now possible and necessary.

We have considered a large sample of extragalactic H~II regions which
provides an extensive coverage of a very wide metallicity range
(almost a factor of 100), and includes galaxies with a variety of
morphological types and luminosities. Such a sample is in many
respects much more homogeneous than any sample of galactic H~II
regions. All the H~II regions observed are large (diameter D $>$ 50
pc), tenuous (n$_e$ $<$ 500 cm$^{-3}$ as derived from the [S II]
I(6717)/I(6731) ratio) and must be ionized by large OB associations.

Here, we limit our analysis to data published as of February 1992. (A
more complete investigation, including the discussion of data
published as of December 1999, is in progress and will be completed
soon; Lenzuni and Panagia 2000, in preparation). Thus, our sample 
currently includes 287
H~II regions in 46 spiral and irregular galaxies with positive
detections of the [O II] lines at 3726 and 3729 \AA~ (usually
unresolved), of the [O III] lines at 4959 and 5007 \AA~, and of at
least one of the He I lines. Additional observations were also
collected for 87 ``Blue Compact Galaxies'' (BCG's). None of these
objects is resolved into individual H~II regions, the observations being
relative to the entire galaxy or, possibly, to its central, brightest
parts. These galaxies appear to be undergoing a stage characterized by
a collective mode of star-formation. Their spectra are heavily
dominated by H~II region-like emission, hence they can be treated for
our purposes as giant, isolated, extragalactic H~II regions. 

\section{Analysis and Discussion}
% \subsection{The Helium Enrichment Curve}

Ionized helium abundances are shown in Figure 1 as a function of
oxygen abundances, for all of the H~II regions in spiral and irregular 
galaxies and the Blue Compact
Galaxies of our sample. 
The long baseline in metallicity offers a unique opportunity to
constrain {\it both } the abundance of primordial helium Y$_P$ {\it
and} to the \dYdZ gradient, thus fully determining the ``helium
enrichment curve'' (HEC) which relates the total abundance of helium
to the abundance of oxygen. Considering that evolutionary effects
always decrease the  He$^+$/H$^+$ ratio because the
aging of a stellar cluster results in a softening of the ionizing
radiation field,  and that young clusters are observationally favoured
because they are intrinsically brighter than older clusters, the HEC
can be derived by determining the upper envelope of the distribution
shown in Figure 1. Among the class of curves $Y = Y_0 + \Delta
Y/\Delta Z \times Z$ the best fit to the upper envelope of the
observations is obtained for $Y_0$ = 0.243 and 
(\dYdZ)$_{\odot}$ = 3.2 (see dashed curve in Fig.~1).
 This relation is consistent with the observational results of Pagel
\etal~ (1992) as well as with Maeder's (1992) theoretical models. 
An inspection to Figure 1 reveals that the observed He$^+$/H$^+$ ratio
appears to be almost constant up to solar O abundance
(log(O/H)$_\odot+12\simeq$8.8) and then it declines rather quickly for higher
metallicities.  This is a clear sign that He is progessively less
ionized as the O abundance increases, and implies that the mean
radiation temperature of the ionizing stars becomes lower than about
38,000~K around log(O/H)$\simeq8.5$. 

\begin{figure}
\centerline{
 \psfig{figure=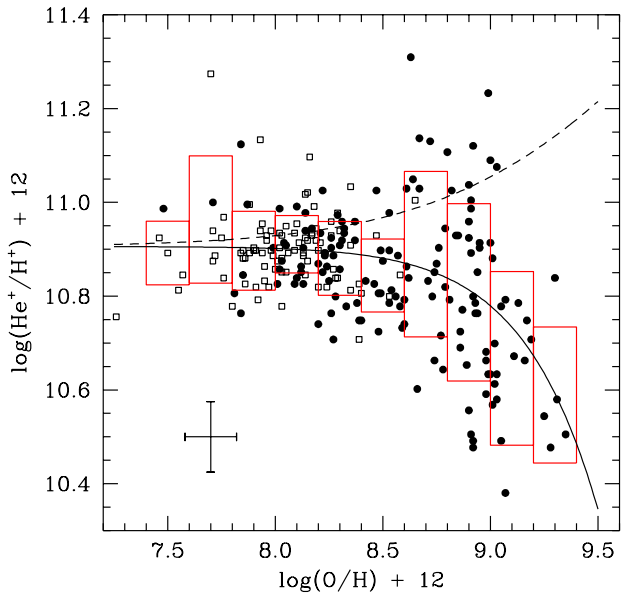,width=9cm,height=9cm}
}
%\vskip-.3cm
\caption{ He$^+$/H$^+$ ratios (by number) as a function of the 
O~abundances for the 287 H~II regions in 46 Spiral and Irregular
galaxies (filled symbols) and for the 87 Blue Compact Galaxies (open
squares) of our sample. The cross in the lower left corner represents
the typical uncertainty of the data. The boxes correspond the
$\pm1\sigma$ deviations from the averages calculated within 0.2 dex
intervals in log(O/H). The dashed line represents the helium
enrichment curve (HEC) and the solid line is a model calculation in
which the He ionization decreases exponentially with increasing 
O~abundance.} 
\end{figure}

% \vskip-.8cm
We find that mechanisms through which metallicity acts ``directly'' on
the radiation field of the ionizing star cluster are not enough to explain the
observed gradient of the He ionization fraction with O 
abundance. Most of the effect appears instead to be due to
``indirect'' mechanisms, \ie a marked deficiency of hot stars
with increasing metal abundances. There are at least three possible
scenarios to explain this fact:\\ 
1. The IMF slope becomes steeper for higher metallicities.\\
2. The IMF upper cutoff moves to lower masses for higher 
metallicities.\\
3. The most massive stars become progressively unable to provide
ionizing radiation, either because at high metallicities the remnant
of their pre-MS cocoons remains optically thick over most of a star's
lifetime, or because pulsational instabilities prevent the most massive
stars from reaching their expected ZAMS surface conditions. \\
Our model calculations show that varying only the slope of the IMF,
\ie point (1), does not give a satisfactory fit to the data because
the resulting ionization decline would be too shallow. On the other hand,
point (2), \ie a systematic variation of the IMF upper mass cut-off
with metal abundance ($m_U \propto$ Z$^{-\beta}$, $\beta > 0$) can
reproduce the observed trend of the He ionization, with $m_{U\odot}$ =
48 \Msun and $\beta$ = 0.60. Point (3) could also account for the
observations provided that the invoked effects are indeed capable to
produce the sharp decline of He ionization as observed. 

From the observational point of view there are no direct studies to
conclusively discriminate between points (2) and (3). Observations of
massive stars near the Galactic Center, such as the Pistol star, 
the Sickle and the Quintuplet clusters (\eg Figer \etal~ 1999 and
references therein) seem to favor the third possibility, because they
are so bright ($log(L/L_\odot)>6$) that they must be quite massive. 
On the other hand, one may argue that those clusters are so close to
the Galactic Center that tidal forces may drastically affect the
dynamical processes that lead to the formation of stars and, therefore,
they may not be representative of normal situations. The ideal
investigation to clarify this issue should include nebular and stellar
spectroscopy of a large sample of HII regions in galaxies which
display marked effects  of incomplete He ionization, such as M51 or 
M83.  Another discriminant between hypotheses (2) and (3) is that if
a lowering of the IMF upper cutoff is {\it the } explanation (\ie
point (2)), then the frequency of Wolf-Rayet stars relative to early
type stars is expected to be abnormally low in the high-Z regions
because in this case the reduced ionization is entirely due to the
lack of truly massive stars that are expected to end up as WR stars in
their final stages of evolution.

\end{document}